\documentclass{iau} 

\newcommand{\aap}{{\it A\&A}}

\newcommand{\aj}{{\it AJ}}
\newcommand{\apj}{{\it ApJ}}
\newcommand{\apjl}{{\it ApJL}} 
\newcommand{\apjs}{{\it ApJS}}

\newcommand{\mnras}{{\it MNRAS}}

\newcommand{\pasp}{{\it PASP}}

\usepackage{graphicx}

\title[Searching for Misidentified Massive Stars in Algorithmically-Selected Quasar Catalogs] 
{There Are (super)Giants in the Sky: \\ Searching for Misidentified Massive Stars in Algorithmically-Selected Quasar Catalogs}

\author[Trevor Z. Dorn-Wallenstein \& Emily Levesque]   
{Trevor Z. Dorn-Wallenstein
 \and Emily Levesque}

\affiliation{Astronomy Department, University of Washington, \\ Physics and Astronomy Building, 3910 15th Ave NE, Seattle, WA 98105, USA \\ email: {\tt tzdw@uw.edu, emsque@uw.edu} }

\pubyear{2016}
\volume{329}  
\jname{The Lives and Death-Throes of Massive Stars}
\editors{J.J. Eldridge, ed.}
\begin{document}

\maketitle

\begin{abstract}

Thanks to incredible advances in instrumentation, surveys like the Sloan Digital Sky Survey have been able to find and catalog billions of objects, ranging from local M dwarfs to distant quasars. Machine learning algorithms have greatly aided in the effort to classify these objects; however, there are regimes where these algorithms fail, where interesting oddities may be found. We present here an X-ray bright quasar misidentified as a red supergiant/X-ray binary, and a subsequent search of the SDSS quasar catalog for X-ray bright stars misidentified as quasars. 

\end{abstract}

\firstsection 
\section{Introduction/Overview}
 
\subsection{Red Supergiant X-ray Binaries}

Over the past decade, many exotic close binary systems with supergiant components have been discovered. Systems like NGC 300 X-1 --- a Wolf-Rayet/black hole X-ray binary (\cite[Crowther \etal \ 2010]{crowther10}) --- and SN2010da --- a sgB[e]/neutron star X-ray binary (\cite[Villar \etal \ 2016]{villar16}) --- are two such examples of a coupling between a massive star in a short-lived evolutionary phase and a compact stellar remnant. Interestingly, no X-ray binaries with confirmed red supergiant (RSG) counterparts have been discovered (RSGs have been proposed as the candidate donor star for a few Ultraluminous X-ray Sources, see \cite[Heida \etal \ 2016]{heida16}). This may be partially explained by the rarity of RSGs; however, though rare, RSGs are both longer-lived and more common (due to the smaller --- 10 - 25 $M_\odot$ --- initial masses of their zero age main sequence progenitors) than most other evolved massive stars. 

RSG X-ray binaries, if they exist, offer a view into an interesting edge case of accretion; their extended envelopes and strong winds ($M \sim 10^{-4}\:M_\odot$ yr$^{-1}$, \cite[van Loon \etal \ 2005]{vanloon05}) could allow for accretion from both the wind and Roche-Lobe Overflow in an environment continually enriched with dust produced by the RSG. RSG X-ray binaries are also the immediate progenitors of Thorne-\.{Z}ytkow Objects --- stars with embedded neutron star cores (\cite[Thorne \& \.Zytkow \ 1975]{thorne75}) --- assuming the neutron star plunges into the RSG as it expands (\cite[Taam \etal \ 1978]{taam78}). 

\subsection{J0045+41}

To search for RSG X-ray binaries, we used the photometry of the Local Group Galaxy Survey (LGGS, \cite[Massey \& Olsen 2003, Massey \etal \ 2006, 2007]{massey03,massey06,massey07}), which covers M31, M33, the Magellanic Clouds and 7 dwarf galaxies in the Local Group. Following \cite{massey98} to find RSGs among the nearly-identical foreground dwarfs, we cross-referenced the positions of the LGGS RSGs with the {\it Chandra} Source Catalog (CSC, \cite[Evans \etal \ 2010]{evans10}), and found one RSG coincident with an X-ray source.

LGGS J004527.30+413254.3 (J0045+41 hereafter) is a bright ($V \approx 19.9$) object of previously-unknown nature in the disk of M31. \cite{vilardell06} classify J0045+41 as an eclipsing binary with a period of $\sim 76$ days. J0045+41 was also observed with the Palomar Transient Factory (PTF); the $g$-band lightcurve shows evidence for a $\sim650$ day period. On the other hand \cite{kim07} identify J0045+41 as a globular cluster, and it has been included in catalogs of M31 globular clusters as recently as 2014 (\cite[Wang \etal \ 2014]{wang14}). The LGGS photometry is consistent with the color and brightness of a RSG. Indeed, following \cite{levesque06}, we found that, as an RSG, J0045+41 would have an effective temperature of $\sim$3500 K and bolometric magnitude of -6.67, consistent with a 12-15 M$_{\odot}$ RSG. However, a complete SED fit to photometry from the Panchromatic Hubble Andromeda Treasury (PHAT, \cite[Dalcanton \etal 2012]{dalcanton12}) using the Bayesian Extinction and Stellar Tool (BEAST, \cite[Gordon \etal \ 2016]{gordon16}) yields an unphysical result of 300 M$_{\odot}$, 10$^{\rm 5}$ K star, extincted by $A_V\sim$4 magnitudes. Furthermore, the object appears extended in the PHAT images (though its radial profile appears similar to that of other nearby stars). 

J0045+41 is separated by $\sim1.18^{\prime\prime}$ from an X-ray source. The source, CXO J004527.3\\+413255, is bright ($F_X = 1.98\times10^{-13}$ erg s$^{-1}$ cm$^{-2}$) and hard; fitting a spectrum obtained by Williams et al. (in prep) yields a power law with $\Gamma\sim1.5$. The best-fit neutral hydrogen column density is $1.7\times10^{21}$ cm$^{-2}$, which corresponds to $A_V\sim1$. 

\subsubsection{Observations and Data Reduction}

The apparent periodicity of J0045+41 and its apparent association with a hard and unabsorbed X-ray source prompted us to obtain follow-up spectroscopic observations to determine the true nature of this object. We obtained a longslit spectrum of J0045+41 using the Gemini Multi-Object Spectrograph (GMOS) on Gemini North. Four 875 second exposures were taken 2016 July 5 using the \texttt{B600} grating centered on 5000 \AA, and four 600 second exposures were taken 2016 July 9 using the \texttt{R400} grating centered on 7000 \AA, with a blocking filter to remove 2$^{nd}$-order diffraction. Due to the gaps between GMOS's three CCDs, two of each set of exposures were offset by +50 \AA. The data were reduced using the standard \texttt{gemini} \texttt{IRAF} package. The final reduced spectrum has continuous signal from $\sim$4000 to $\sim$9100 \AA \ at a resolution of $R \sim 1688\:({\rm blue})/1918\:({\rm red})$.

\subsubsection{Spectrum and Redshift Determination}

\begin{figure}
  \centering
    \includegraphics[width=0.9\textwidth]{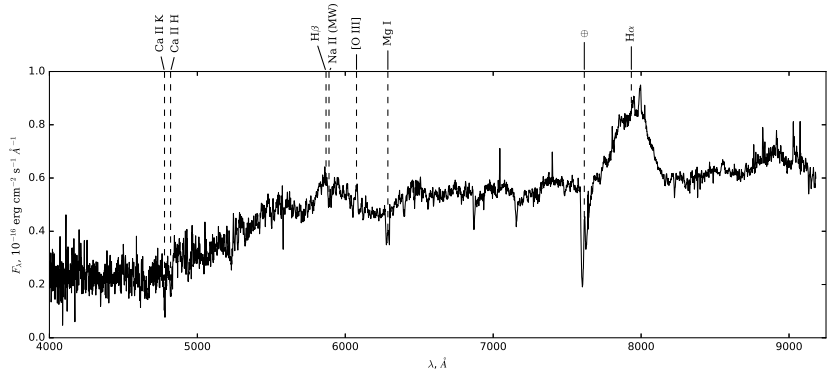}
  \caption{GMOS spectrum of J0045+41 with all identified lines labeled. The Na II feature is intrinsic to the Milky Way/M31, and the $\oplus$ line is telluric.}\label{fig:spectrum}
\end{figure}

The spectrum (Figure \ref{fig:spectrum}) shows that J0045+41 is a quasar at $z\approx0.21$ (measured with H$\alpha$, H$\beta$, [OIII]$\lambda5007$ and Ca II H and K). While a false positive, this quasar is quite interesting on its own. The (low-significance) detections of periodicity on short timescales by multiple sources are difficult to explain. Furthermore, mistaking a blue quasar for a red star would imply a high reddening along the line of sight, which is consistent with a sightline through M31; however, the low H column density implied by the X-ray counterpart's spectrum and our inability to achieve a satisfactory fit to the optical spectrum by reddening the (redshifted) quasar template spectrum from \cite{vandenberk01} indicate that J0045+41 belongs to a small and intriguing class of intrinsically red quasars, observed through a low-extinction region of the ISM in M31 (\cite[Richards \etal \ 2003]{richards03}).

\section{Stars in the SDSS Quasar Catalog}

If a quasar can be misidentified as a RSG, are there red stars --- especially X-ray bright stars --- in already-existing quasar catalogs? To answer this question, we turned to the quasar catalog of the Sloan Digital Sky Survey (SDSS, \cite[York \etal \ 2000]{york00}). 

\subsection{Sample Selection}

We selected all SDSS objects automatically tagged as quasars that were within 0.2 magnitudes of J0045+41 in $g-r$ vs. $r-i$ vs. $i-z$ color-space --- J0045+41 is too faint in $u$ to utilize $u-g$ --- and ignored any warning flags to avoid throwing out interesting objects that were not easily identified by the SDSS algorithm. 1098 objects in this sample had associated spectroscopic observations. Interestingly, many of the spectroscopically-determined redshifts were unbelievably small or even negative, implying that these objects are in a regime of color-space where classification algorithms may fail. Indeed, on visual inspection of these spectra, many of them are stellar.

\subsection{Stars}

\begin{figure}
  \centering
    \includegraphics[width=0.9\textwidth]{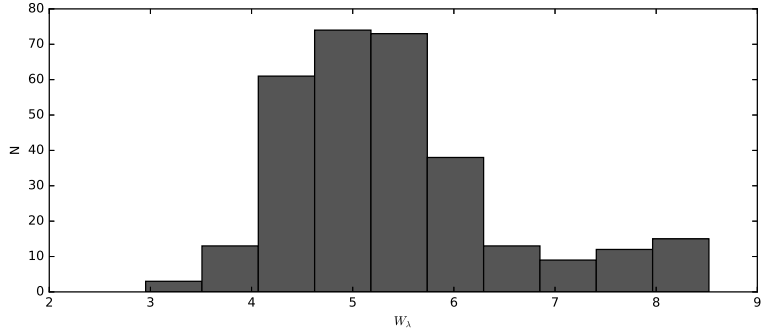}
  \caption{Distribution of measured equivalent widths for confirmed stars.}\label{fig:ew_hist}
\end{figure}

We used \texttt{emcee}, a Python implementation of Markov Chain Monte Carlo (MCMC) by \cite{formanmackey13}, to fit Gaussian profiles to the Ca II triplet ($\lambda = 8498,\: 8542,\: {\rm and}\: 8662$ \AA), which we use to identify stars. The posterior distributions of the parameters allow us to determine if the triplet is well fit, and estimate errors for each parameter. Because the relative centroids and strengths of the lines are fixed, a good fit guarantees the lines are actually detected, while simultaneously measuring --- with accurate errors --- the radial velocity and equivalent width ($W_\lambda$) of the triplet. After a follow-up inspection by eye of spectra that were noisy or missing data, we find 344 confirmed cool stars, representing $\sim$31\% of the total sample. Figure \ref{fig:ew_hist} shows the distribution of $W_\lambda$ for $W_{\lambda}/\sigma_{W_\lambda} > 1$. We follow \cite{jennings16} to estimate luminosity from $W_\lambda$,  and find that most stars are dwarfs ($W_{\lambda} \lesssim 6.5$ \AA) but $\sim$40 stars have larger equivalent widths indicating they are likely giants or supergiants (the relationship is dependent on effective temperature and metallicity, so these labels are approximate).

\section{Discussion and Future Work}

This result demonstrates that, when looking for rare objects like RSG X-ray binaries, it is important to look in unlikely places; e.g., a red and X-ray bright star may be confused for a quasar if the classification algorithm mistakes the continuum between the TiO bands in a RSG spectrum for an emission line. The fact that some of the color-space containing M-dwarfs --- by far the most common type of star --- is a regime where classification algorithms fail underlines the importance of improving on these algorithms until they perform as well as the human eye. Indeed, many of these stars were previously identified (see \cite[West \etal \ 2011]{west11}), but are still listed as quasars on the SDSS online data portal. 

Future work will focus on improving our star-finding algorithm to use alternate spectral features when the Ca II triplet is missing or obscured by noise, and on finding which areas of color-space contain significant numbers of these misidentified stars, with the goal of finding RSG X-ray binaries as well as improving our knowledge of where classification algorithms fail.


\begin{thebibliography}{}

\bibitem[Crowther \etal (2010)]{crowther10}{Crowther, P.~A., Barnard, R., Carpano, S., \etal} 2010, \mnras, 403, L41 

\bibitem[Dalcanton \etal (2012)]{dalcanton12} Dalcanton, J.~J., Williams, B.~F., Lang, D., et al.\ 2012, \apjs, 200, 18 

\bibitem[Davidsen \etal (1977)]{davidsen77} Davidsen, A., Malina, R., \& Bowyer, S.\ 1977, \apj, 211, 866 

\bibitem[Evans \etal (2010)]{evans10} Evans, I.~N., Primini, F.~A., Glotfelty, K.~J., et al.\ 2010, \apjs, 189, 37-82 


\bibitem[Foreman-Mackey \etal \ (2013)]{formanmackey13} Foreman-Mackey, D., Hogg, D.~W., Lang, D., \& Goodman, J.\ 2013, \pasp, 125, 306 


\bibitem[Gordon \etal (2016)]{gordon16} Gordon, K.~D., Fouesneau, M., Arab, H., et al.\ 2016, \apj, 826, 104 

\bibitem[Heida \etal (2016)]{heida16}{Heida, M., Jonker, P.~G., Torres, M.~A.~P., \etal} 2016, \mnras, 459, 771 

\bibitem[Jennings \& Levesque \ (2016)]{jennings16} Jennings, J., \& Levesque, E.~M.\ 2016, \apj, 821, 131 

\bibitem[Kim \etal \ (2007)]{kim07} Kim, S.~C., Lee, M.~G., Geisler, D., et al.\ 2007, \aj, 134, 706 

\bibitem[Levesque \etal \ (2006)]{levesque06} Levesque, E.~M., Massey, P., Olsen, K.~A.~G., et al.\ 2006, \apj, 645, 1102 

\bibitem[Massey (1998)]{massey98} Massey, P.\ 1998, \apj, 501, 153 

\bibitem[Massey \& Olsen(2003)]{massey03} Massey, P., \& Olsen, K.~A.~G.\ 2003, \aj, 126, 2867 

\bibitem[Massey \etal (2006)]{massey06} Massey, P., Olsen, K.~A.~G., Hodge, P.~W., et al.\ 2006, \aj, 131, 2478 

\bibitem[Massey \etal (2007)]{massey07} Massey, P., Olsen, K.~A.~G., Hodge, P.~W., et al.\ 2007, \aj, 133, 2393 

\bibitem[Richards \etal \ (2003)]{richards03} Richards, G.~T., Hall, P.~B., Vanden Berk, D.~E., et al.\ 2003, \aj, 126, 1131 

\bibitem[Taam \etal \ (1978)]{taam78} Taam, R.~E., Bodenheimer, P., \& Ostriker, J.~P.\ 1978, \apj, 222, 269 

\bibitem[Thorne \& \.Zytkow (1975)]{thorne75} Thorne, K.~S., \& \.Zytkow, A.~N.\ 1975, \apjl, 199, L19 

\bibitem[Vanden Berk \etal \ (2001)]{vandenberk01} Vanden Berk, D.~E., Richards, G.~T., Bauer, A., et al.\ 2001, \aj, 122, 549 

\bibitem[van Loon \etal (2005)]{vanloon05} van Loon, J.~T., Cioni, M.-R.~L., Zijlstra, A.~A., \& Loup, C.\ 2005, \aap, 438, 273 

\bibitem[Vilardell \etal \ (2006)]{vilardell06} Vilardell, F., Ribas, I., \& Jordi, C.\ 2006, \aap, 459, 321 

\bibitem[Villar \etal (2016)]{villar16}{Villar, V.~A., Berger, E., Chornock, R., \etal} 2016, \apj, 830, 11 

\bibitem[Wang \etal (2014)]{wang14} Wang, S., Ma, J., Wu, Z., \& Zhou, X.\ 2014, \aj, 148, 4 

\bibitem[West \etal (2011)]{west11} West, A.~A., Morgan, D.~P., Bochanski, J.~J., et al.\ 2011, \aj, 141, 97 

\bibitem[York \etal \ (2000)]{york00} York, D.~G., Adelman, J., Anderson, J.~E., Jr., et al.\ 2000, \aj, 120, 1579

\end{thebibliography}
\end{document}